\documentclass[prd,twocolumn,showpacs,nofootinbib,floatfix]{revtex4}

\ifx\pdfoutput\undefined
   \usepackage[dvips]{graphicx}
   \else
   \usepackage[pdftex]{graphicx}
   \fi
   \usepackage{epstopdf}

\usepackage{amsfonts}
\usepackage{amsmath}
\usepackage{amssymb}
\usepackage{bm}
\usepackage{dcolumn}
\usepackage{epsfig}
\usepackage{graphicx}
\usepackage{graphics}
\usepackage[latin1]{inputenc}
\usepackage{latexsym}
\usepackage{rotating}
\usepackage{hyperref}
\usepackage[usenames]{color}

\newcommand\be{\begin{equation}}
\newcommand\ba{\begin{eqnarray}}
\newcommand\ee{\end{equation}}
\newcommand\ea{\end{eqnarray}}

\newcommand{\ppN}{{\mbox{\tiny ppN}}}
\newcommand{\ppE}{{\mbox{\tiny ppE}}}
\newcommand{\ppK}{{\mbox{\tiny ppK}}}
\newcommand{\GR}{{\mbox{\tiny GR}}}

\newcommand{\SPA}{{\mbox{\tiny SPA}}}

\begin{document}
\title{A Rosetta Stone for Parameterized Tests of Gravity}

\author{Laura Sampson}
\affiliation{Department of Physics, Montana State University, Bozeman, MT 59717, USA.}

\author{Nicol\'as Yunes}
\affiliation{Department of Physics, Montana State University, Bozeman, MT 59717, USA.}

\author{Neil Cornish}
\affiliation{Department of Physics, Montana State University, Bozeman, MT 59717, USA.}

\date{\today}


\begin{abstract}

Several model-independent parameterizations of deviations from General Relativity have been developed to test Einstein's theory. Although these different parameterizations were developed for different gravitational observables, they ultimately all test the same underlying physics. In this paper, we develop connections between the parameterized post-Newtonian, parameterized post-Keplerian, and the parameterized post-Einsteinian frameworks, developed to carry out tests of General Relativity with Solar System, binary pulsar, and gravitational wave observations respectively. These connections allow us to use knowledge gained from one framework to inform and guide tests using the others. Relating these parameterizations and combining the results from each approach strengthens our tests of General Relativity. 
 
\end{abstract}


\maketitle

\section{Introduction}

{\renewcommand{\arraystretch}{2.0}

Since its publication in 1915, Einstein's theory of General Relativity (GR) has been tested extensively~\cite{lrr-2006-3} . To-date, it has survived every challenge. Beginning with the early tests of very weak-field gravity in our solar system, through to the most extreme laboratories of gravitational physics that we can yet access - binary pulsars - GR has had its predictions vindicated with extreme precision. There is, so far, no experimental evidence for GR being incorrect on small scales (i.e.~solar system and compact binary systems), though
flat galaxy rotation curves~\cite{1983ApJ...270..365M, 1983ApJ...270..371M} and the accelerated expansion of the Universe~\cite{Copeland:2006wr} have been put forward as evidence for deviation on much larger scales. On small scales, and for weak fields, the constraints on deviations from GR are so strong that we can say that alternative theories of gravity have effectively been ruled out in this regime. 

But it is not true that we have effectively ruled out all alternative theories of gravity in all regimes. Aside from being an impossible endeavor, given our continued ingenuity in developing new theories of gravity, definitively ruling out all challengers to GR will require tests in the strong-field, dynamical gravitational regime. This is a regime that we have not yet been able to access.  

The gravitational compactness of a system is defined as  \cite{Yunes:2013dva}
\be
\mathcal{C} = \frac{G}{c^2} \frac{M}{R},
\ee
where $M$ is the characteristic mass of the system, and $R$ is the characteristic length scale associated with gravitational radiation. The characteristic velocity of a system, $\mathcal{V}$, is a measure of the rate of change of the gravitational field. The weak field, then, is defined as the regime in which both $\mathcal{C}$ and $\mathcal{V}$ are very small, i.e. $\mathcal{C}\ll1$ \emph{and} $\mathcal{V}\ll1$. The strong-field, dynamical regime is the regime in which \emph{neither} of these conditions are met and a lowest-order perturbative analysis of the gravitational field equations does not suffice. 

We have not yet been able to measure systems in the strong-field. Even for the double pulsar binary system (PSR J0737-3039)  \cite{Kramer:2006nb} , which boasts the strongest gravitational fields we have yet been able to directly probe, the compactness of the system is only $\mathcal{C} \sim 6\times10^{-6}$, with characteristic velocity of $\mathcal{V} \sim 2\times10^{-3} $. In contrast, compact binary coalescences will reach both compactness and velocity close to unity. Thus while binary pulsar data can teach us about strong gravitational fields, they are not dynamical, strong-field systems by our definition. Our lack of data
in this regime means that there is still room left for alternative theories of gravity that do not predict outcomes strongly different from GR except in very strong-field, dynamical systems. 

Most alternative gravity theories, and there are many \cite{Yunes:2013dva}, involve additional parameters in their field equations, such as coupling constants, or the mass of extra fields. These parameters are inherent to the theory, but their values are unknown and must be measured. Clearly, the parameters for each theory could be constrained via experiment, for instance by predicting the outcome of a given experiment for a given alternative theory of gravity. Although this is an obvious way forward, the work involved in making testable predictions for every known theory of gravity is enormous - even for the limited subset of theories that we have so far discovered. It is therefore important to develop methods for testing multiple theories of gravity simultaneously. It is with this goal in mind that the various generically parameterized models of non-GR gravity have been developed.

The parameterized post-Newtonian (ppN) formalism, developed by Will and Nordtvedt \cite{Nordtvedt:1968qs,1972ApJ...177..757W,1972ApJ...177..775N,1971ApJ...163..611W}, parameterizes the space-time metric in a way that captures many different types of gravitational theories. It is built on an expansion about Minkowski space, and is accurate in the weak-field regime, and on scales that are sufficiently small to ensure the accuracy of a linear perturbation. The ppN framework was designed for, and is thus ideally suited to, tests we can perform in our own solar system.

 In contrast, the parameterized post-Keplerian (ppK) formalism was designed to allow us to perform tests of GR with pulsar timing data \cite{Freire:2010sv,Ruggiero:2005dg,Iorio:2004ug}. Developed in its current form primarily by Damour, Deruelle, and Taylor \cite{1992PhRvD..45.1840D, Damour1985}, it gives a timing formula for the arrival of pulses at Earth that builds on an expansion on Kepler's laws. The timing formula phenomenologically takes into account many different gravitational effects, both in the binary motion of the components of a pulsar system, and on the travel of electromagnetic (EM) radiation away from such a system towards Earth. Measuring these parameterized effects allows us to test the nature of the underlying theory of gravity that describes our universe in regimes where the gravitational field is as strong as that induced by binary pulsars.
 
Adding to our knowledge from solar system experiments and measurements of the pulses from binary pulsar systems, in the near future we will have yet another tool for testing the fundamental nature of gravity - gravitational waves (GWs). The Fourier amplitude and phase of GWs from inspiraling compact objects have been predicted to extreme precision within GR \cite{lrr-2006-4}, but not within alternative theories. To solve this problem, and also to allow a means of detecting or constraining un-modeled deviations from GR, Yunes and Pretorious ~\cite{Yunes:2009ke} developed the parameterized post-Einsteinian (ppE) family of waveform templates. The ppE formalism has since been extended by Chatziioannou,  Yunes, and Cornish~\cite{Chatziioannou:2012rf} to describe the multiple polarization states that generically appear in extensions to GR. The parameters in the ppE model can be constrained by GW detections, and these constraints in turn will either build our confidence that GR is the correct theory of gravity, or indicate that we must deepen our knowledge of other possible theories in order to explain our observations.
 
In addition to these three parameterizations, there are also the parameterized post-Friedmann (ppF)~\cite{Hu:2007pj,PhysRevD.87.024015} approaches, which add parameters to the linearized Einstein field equations in cosmological perturbation theory to parameterize various gravity models. This allows for the use of cosmological data to constrain the nature of gravity. Because the ppF formalisms are applicable over extremely large scales, and usually
involve mechanisms to suppress deviations on the smaller scales probed by the other parameterizations, we will not address them in this paper.
 
The ppN and ppK formalisms have already been used successfully to constrain GR in relatively weak fields. New tests of GR using GWs will allow us to probe gravitational fields that are much stronger and more dynamical than anything we have had access to before. Yet we know that the underlying theory of gravity describing these types of systems is the same as that which governs the physics of our solar system and of pulsars. It would therefore be sub-optimal to proceed as if we have no knowledge of the constraints already placed on alternative theories of gravity when analyzing new data. 

The purpose of this paper is to draw connections between the ppE, ppN, and ppK formalisms, so that measurements that constrain parameters in one can be used to enhance our knowledge of constraints in the other two. Knowing the connection between the ppN, ppK, and ppE parameters will allow us to use known constraints from the solar system and binary pulsars to inform our search for deviations from GR in GWs. For instance, they can be used to construct a well-informed prior when conducting a Bayesian model-selection analysis~\cite{Cornish:2011ys,Sampson:2013lpa}. Going in the other direction, constraining ppE parameters using GWs could allow us to place stronger constraints on ppN and ppK parameters than we have been able to achieve using solar system and binary pulsar tests. 

The main result of this paper is the mathematical mapping between the three different parameterizations. First, we find the mapping between ppN and ppE parameters. In doing so, we discover that ppN modifications to the binding energy generically lead to 1 post-Newtonian (PN) corrections to the GW phase, if and only if the ppN parameters $\beta_{\ppN}, \alpha_1^\ppN,$ or $\alpha_2^\ppN$ (see Sec.~\ref{sec:zoo} for definitions) are modified. Most alternative theories studied thus far do not modify these parameters, which is perhaps why leading-order GR deviations that enter at 1PN order have not yet been found. Second, we find the mapping between ppK and ppN parameters. We find that although the Shapiro shape parameter is not modified by ppN corrections, the range parameter, the redshift parameter and the pericenter rate of change parameter are modified and directly related. Third, we find the mapping between ppK and ppE parameters by investigating ppE modifications to the conservative dynamics. 

With these mappings at hand, we can then investigate how current bounds on ppN and ppK parameters already constrain ppE parameters. We find that binary pulsar constraints on ppK parameters already stringently bound ppE parameters in a certain weak-field regime of ppE parameter space, ie.~when the ppE phase exponent parameter $b < 4$. In this regime, these bounds are stronger than future projected bounds with GWs~\cite{Cornish:2011ys}. We also find that Solar System constraints on the ppN parameters already stringently bound ppE parameters at 1PN order. Although this constraint holds only at a single point of ppE parameter space, they beat projected bounds from GWs by a factor of a few. Multiple GW observations, or one especially bright signal, will however be able to provide a more stringent bound than that inferred from bounds on ppN parameters from Solar System observations.

The rest of this paper is arranged as follows. In Sec. \ref{sec:zoo}, we give a brief introduction of the ppN, ppK, and ppE parameterizations. In Sec \ref{sec:ppNppE}, we find the correspondence between the ppN metric parameters and the ppE GW parameters. In Sec \ref{sec:ppNppK}, we find the correspondence between ppN and the pulsar timing parameters of ppK. Next, in Sec \ref{sec:ppKppE}, we show the connection between the orbital decay parameter of ppK, and the phase parameters of ppE. Finally, in Sec \ref{sec:Constraints}, we discuss current constraints on the various parameters, and in Sec \ref{sec:Conc}, we conclude and point to future research.
 
Throughout this paper, we use units in which $G = c = 1$. In addition, Greek indices refer to space-time coordinates, and Latin indices refer only to spatial coordinates.
 
\section{parameterized post-taxonomy \label{sec:zoo}}

\subsection{Parameterized post-Newtonian }

PN formalism expands Einstein's equations beginning with the lowest-order deviations from Newtonian gravity. The small expansion parameter is typically the characteristic velocity of the system, which must be small in comparison to the speed of light. The formalism is valid only for weak fields, in systems composed of objects that travel slowly when compared to the speed of light - for instance, in our own solar system.

The ppN formalism, due primarily to Nordtvedt and Will  \cite{Nordtvedt:1968qs,1972ApJ...177..757W,1971ApJ...163..611W}, is a first-order PN framework to parameterize a large class of gravitational theories. It is useful for testing GR in the weak-field, and was developed with solar-system experiments in mind. The following outline of the ppN formalism closely follows Chapter 4 of \cite{ Will:1993ns}.

In order for alternative theories of gravity to be studied in the ppN framework, they must be metric theories - i.e. theories in which there exists a symmetric tensor called the metric that governs proper distances and proper times, and in which matter and fields respond when acted upon by gravity via the equation
\be
\nabla_\nu T^{\mu \nu} = 0,
\ee
where $T^{\mu\nu}$ is the stress-energy tensor for all matter and non-gravitational fields, and $\nabla_{\nu}$ is the covariant derivative with respect to the metric. This requirement is equivalent to the statement that all theories of gravity in the ppN formalism must satisfy the Einstein equivalence principle, which states that, in a freely falling reference frame, all physical laws behave as if gravity were absent. In other words, there are no local experiments that one can perform to differentiate between a uniformly accelerated reference frame, and one that is freely falling with respect to the local gravitational field \cite{ Will:1993ns}.

To generate the ppN formalism, we must begin with some book-keeping that allows us to keep track of the various ``smallness'' parameters used in the expansion. The first assumption is that the gravitational fields are \emph{weak}. In this context, this means that the classical, Newtonian potential of all rest masses in the system is small:
\be
 U(t,\mathbf{x}) \equiv \int\frac{\rho(t,\mathbf{x'})}{|\mathbf{x}-\mathbf{x'}|} d\mathbf{x'} \ll 1,
\ee
where $\rho(t,\mathbf{x'})$ is the rest-mass density at $(t, \mathbf{x'})$. 

In Newtonian gravity, the velocities of bodies in orbit around each other are related via the virial theorem to $U$ through $v^2 \lesssim U$. These quantities are both considered to be of second-order, whereas quantities with a single power of velocity are first-order. For example, $v U$ is a perturbative quantity of $\mathcal{O}(3)$. 

With this book-keeping in hand, and recalling that in this formalism all gravitational theories are governed by a space-time metric, we can proceed to construct a general, ppN metric. As the ppN formalism is based on a perturbation of Newtonian gravity, we know that the ppN metric must reduce to the Minkowski metric at lowest order. Thus the most general metric one could construct would begin with the Minkowski metric, and then add PN metric terms from all possible functionals of matter variables, each multiplied by an arbitrary coefficient that could be set by matching to cosmological conditions. There are, however, an infinite number of these functionals, and so to produce a workable formalism, one typically adopts some restrictions:
\begin{itemize}
\item The metric coefficients should be of Newtonian or (first) PN order, and no higher;
\item Perturbations to the Minkowski metric should go to zero at spatial infinity, so that the metric is asymptotically flat;
\item The metric should be dimensionless;
\item The metric should contain no explicit reference to the spatial origin or the initial moment of time;
\item The metric components $g_{00}, g_{0j}$, and $g_{jk}$ should transform as a scalar, vector, and tensor respectively;
\item The functionals should depend on the rest-mass, energy, pressure, and velocity - not on gradients of these quantities;
\item The functionals should be ``simple.'' 
\end{itemize}

We assume that the theory of gravity of interest can be described by a least-action principle, in which the Lagrangian is defined as 
\be
L = \left(- g_{\mu\nu}\frac{dx^\mu}{dt}\frac{dx^\nu}{dt}\right)^{1/2} =  \left( g_{00} - 2 g_{0j}v^j - - g_{jk} v^{j}v^{k} \right)^{1/2}\,.
\ee
Because the velocity, $v$, is a first-order parameter, in order to keep each of these terms at the same order, we need to know $g_{00}$ to $\mathcal{O}(4)$, $g_{0j}$ to $\mathcal{O}(3)$, and $g_{jk}$ to $\mathcal{O}(2)$. 

\begin{table*}[ht]
\begin{center}
    \begin{tabular}{c|c| p{2.5cm} | p{5cm} }
    \hline\hline
    Parameter & Value in GR & Value in semi-cons. theories & What does it measure?\\ \hline
  $\gamma_{\ppN}$ & 1 & $\gamma_{\ppN}$ & How much space-time curvature is produced by a unit rest mass? \\ 
  $\beta_{\ppN}$ & 1 & $\beta_{\ppN}$ & How much ``nonlinearity'' is there in the superposition law for gravity \\ 
  $\xi$ & 0 & $\xi$ & Are there preferred location effects? \\ 
   $\alpha_1^{\ppN}, \alpha_2^{\ppN}, \alpha_3^{\ppN}$ & 0 & $\alpha_1^{\ppN}, \alpha_2^{\ppN}, 0$ & Are there preferred frame effects? \\ 
    $\zeta_1^{\ppN},\zeta_2^{\ppN},\zeta_3^{\ppN},\zeta_4^{\ppN}$ & 0 & 0 & Violation of mom. conservation?\\  \hline\hline
           \end{tabular}
   
    \caption{ \label{Table:ppNpar}The ten ppN parameters, as well as their physical significance, and their value in GR and in semi-conservative theories, in which energy and momentum are conserved.}
\end{center}
\end{table*}

As an example, let us consider $g_{jk}$.  We have determined that this metric element must transform as a tensor, and contain functionals of the rest-mass, energy, pressure, and velocity that are no higher than second order in our expansion parameters. The only terms that can appear in $g_{jk}$ that satisfy these restrictions, as well as the full list of restrictions above, are $U \delta_{jk}$ and $U_{jk}$ where $U_{jk}$ is given by
\be
U_{jk} \equiv \int \frac{\rho(\mathbf{x'},t) (x - x')_j (x-x')_k}{|\mathbf{x}-\mathbf{x'}|}d^3x'.
\ee
The other metric components can be written in terms of similar functionals, each meeting the requirements described above. There are ten such functionals in total, after many have been eliminated for failing the final, and rather subjective requirement of ``simplicity.'' After the metric is written in terms of these \emph{metric potentials}, we make a choice of gauge and coordinate system that results in the final ppN parameterization.

The metric components are then written, in this coordinate system, as a collection of constants multiplying the metric potentials. For instance, the $g_{jk}$ component becomes
\be
g_{jk} = (1+ 2\gamma_{\ppN} U)\delta_{jk}.
\ee
Here, $\gamma_{\ppN}$ is one of the aforementioned constants. These constants, of which there are ten, are called ``ppN parameters." They are listed in Table \ref{Table:ppNpar} with their physical significance.  In different theories of gravity, these ten parameters take on different values. Measurements of or constraints on these parameters can then either constitute evidence for a non-GR theory of gravity or place bounds on GR deviations.

These ppN parameters have been constrained by many experiments within our solar system, such as lunar laser ranging \cite{WilliamsJG}, gravitational redshift experiments \cite{GodoneA,Dicke}, and the measurement of Earth's tides \cite{Lageos}. The current limits on the ppN parameters, as well as the sources of those limits, are listed in Table \ref{Table:ppNlimits}.
\begin{table}[ht]
\begin{center}
    \begin{tabular}{c|c| c }
    \hline\hline
    Parameter & Effect & Limit\\ \hline
  $\bar{\gamma}_{\ppN}$ & time delay & $2.3 \times 10^{-5}$ \\ 
  $\bar{\beta}_{\ppN}$ & perihelion shift & $3.0\times 10^{-3} $\\ 
  $\alpha_1^{\ppN}$ & orbital polarization & $10^{-4}$\\ 
   $\alpha_2^{\ppN}$ & spin precession & $4\times10^{-7}$\\  \hline\hline
        \end{tabular}
   
    \caption{ \label{Table:ppNlimits}The current experimental constraints on the four ppN parameters that we will consider in this paper, along with the effect used to measure that constraint \cite{lrr-2006-3}. We use the definitions $\bar{\beta}_{\ppN} = \beta_{\ppN} -1,\bar{\gamma}_{\ppN} = \gamma_{\ppN} -1 $}
\end{center}
\end{table}

As stated, all theories of gravity consistent with the ppN formalism obey the Einstein equivalence principle. In addition to these restrictions, one can impose the conservation of total momentum, that is, both momentum and energy. Theories with this conservation law are called ``semi-conservative'', and have only five non-zero ppN parameters. These are $\gamma_{\ppN}, \beta_{\ppN}, \alpha_1^{\ppN}, \alpha_2^{\ppN}$, and $\xi_{\ppN}$. In this paper, in order to be able to compare the ppN results to the post-Keplerian and post-Einsteinian, we restrict ourselves to considering these semi-conservative theories. 

Finally, in order to best understand the bounds that have been placed on this particular type of ppE term, we change variables in many expressions to a system in which all of the ppN parameters are equal to zero in GR. That is, $\{\beta_{\ppN}, \gamma_{\ppN}\} \rightarrow \{\bar{\beta}_{\ppN}, \bar{\gamma}_{\ppN}\}$, where $\bar{\beta}_{\ppN} = \beta_{\ppN} -1,\bar{\gamma}_{\ppN} = \gamma_{\ppN} -1 $.

\subsection{Parameterized post-Keplerian}

We have not only been able to test GR with experiments in our Solar System, but also by analyzing the timing data from pulsars in binary systems \cite{lrr-2003-5}. Pulsars are some of the best clocks in the universe - the arrival time at Earth of their EM pulses can be predicted with extreme precision by fitting the measured arrival times to a timing formula. This formula must include gravitational effects on the emission and travel time of the pulses, as well as non-gravitational effects intrinsic to the pulsar itself. For example, aberration due to the fact that the beam of EM radiation comes from a concentrated point on the star, and not the entire star. The precision and complexity of these systems make them excellent laboratories for testing GR - the simultaneous measurement of several of these effects allows for consistency checks on the theory of gravity used to generate the timing formula.

 Blandford and Teukolsky \cite{1976ApJ...205...580B} derived a timing model that assumed that the components of the binary behaved according to Kepler's laws of planetary motion. The five ``Keplerian'' parameters which enter this timing model are the orbital period, $P_b$, the epoch of periastron passage, $T_0$, the eccentricity, $e$, the longitude of periastron, $\omega$, and the projected semi major axis of the orbit, $x=a \sin\iota / e$, where $a$ is the semi-major axis of the orbit, and $\iota$ is the inclination of the binary, measured from the line of sight to the binary. In addition to the Keplerian parameters, their model allowed for secular drifts of these parameters, as well as one extra parameter, $\gamma_{\ppK}$, to account for special-relativistic time dilation effects. Although Blandford and Teukolsky had intended their model only as a way to measure parameters in GR, the phenomenological approach that they took allowed it to fit timing predictions from other theories as well. The parameters and structure of the timing formulas was theory-independent - it was the functional relationship between the parameters and the masses of the pulsar and its companion that were determined by a given theory of gravity.
 
Later, Epstein \cite{1977ApJ...216...92E} and Haugan \cite{1985ApJ...296....1H} attempted to include the 1PN corrections to the timing formula, which come from the Shapiro time delay and the gravitational redshift due to the mass of the companion, as well as post-Keplerian effects on the orbital motion. The resulting formula was very complicated, and moreover was not theory independent, as it had been calculated within GR. 

Damour and Deruelle \cite{Damour1985} showed that all 1PN corrections to the timing formula could be captured in a simple way that was applicable to many theories of gravity. This work led to their ppK formalism for a pulsar timing formula. This formula includes eight separately measurable post-Keplerian parameters, as well as four post-Keplerian parameters that are not separately measurable. The timing formula including these eight parameters can be written as \cite{1991STIN...9219818D}
\be
t_b - T_0 = F\left[\tau; \{p^K\}; \{p^{pK}\};\{q^{pK}\}\right].
\label{Eq:tmodel}
\ee
Here, $t_{b}$ is the time at which a signal from the pulsar would be detected at the solar-system barycenter if there were no GR effects on its propagation, and $\tau$ is the proper time of the pulsar.

In addition to proper time, the right-hand-side (RHS) of Eq. (\ref{Eq:tmodel}) depends on
\be
\{p^K\} = \{P_b, T_0, e_0, \omega_0, x_0 \},
\ee
the set of Keplerian parameters that describe an elliptical orbit, with $\omega_0$ the initial position of periastron and $x_0$ the initial semi-major axis; 
\be
\{p^{\ppK}\} = \{k, \gamma, \dot{P}_b, r,s,\delta_\theta, \dot{e},\dot{x}\}_{\ppK},
\ee
the set of separately measurable post-Keplerian parameters; and finally 
\be
\{q^{\ppK}\} = \{\delta_r, A,B,D\}_{\ppK},
\ee
the set of not separately measurable post-Keplerian parameters. For a detailed discussion of what exactly each of these parameters is, and how they fit into the timing model, see \cite{1991STIN...9219818D}. 

The right hand side of Eq. (\ref{Eq:tmodel}) can be written schematically as
\be
F(\tau) = D^{-1}\left[\tau + \Delta_R(\tau) + \Delta_E(\tau) + \Delta_S(\tau) + \Delta_A(\tau) \right],
\label{eq:timingschem}
\ee
where each term is due to different effects. In this expression, $\Delta_A$ is called the ``aberration'' delay, and is due to the fact that the pulsar is not simply a radial pulsation, but a rotating beacon. $\Delta_{R}$ is the modulation in arrival time due to the motion of the Earth about the Sun, as well as the orbital motion of the pulsar and its companion, known as the Roemer time delay. $\Delta_{E}$ is the Einstein time delay, or gravitational redshift, caused by the pulsar's binary companion. Finally, $\Delta_{S}$ is the Shapiro time delay, also due to the pulsar's binary companion. 

Each of these effects can be written in terms of a combination of Keplerian and ppK parameters.
\begin{align}
\Delta_R &= x \sin \omega \left[\cos u - e(1+\delta_r) \right] \nonumber \\
&+ x\big[1-e^2(1+\delta_\theta)^{2} \big]^{1/2}\cos \omega \sin u  \\
\Delta_E &= \gamma_{\ppK} \sin u \\
\Delta_S &= - 2 r_{\ppK} \ln \Big\{1 - e \cos u - s_{\ppK} \big[\sin\omega(\cos u -e)\nonumber \\
&  \quad \quad \quad \quad \quad\quad\quad\quad\quad\quad\quad+(1-e^2)^{1/2}\cos\omega\sin u \big] \Big\} \\
\Delta_A &= A \Big\{\sin\big[\omega + A_e(u)\big] + e \sin \omega \Big\} \nonumber \\
&\quad\quad + B\Big\{\cos \big[ \omega + A_e(u)\big] + e \cos \omega \Big\} 
\end{align}
Here, $A_e$ and $\omega$ are functions of $u$, described by
\begin{align}
A_e(u) &= 2 \arctan \left(\left[\frac{1+e}{1-e} \right]^{1/2}\tan\frac{u}{2} \right), \nonumber \\
\omega &= \omega_0 + k_{\ppK} A_e(u),
\end{align}
and finally $u$ is eccentric anomaly, and is a function of proper time, $\tau$. It is defined by solving Kepler's equation
\be
u -e \sin u  = \frac{2\pi}{P_b} (\tau-T_0),
\label{eq:Kepler}
\ee

Later, Damour and Taylor \cite{1991STIN...9219818D} extended the model to include fully nineteen separately measurable ppK parameters, but measurable in theory does not always mean measurable in practice. Of these nineteen parameters, five have  been measured using available pulsar data. These are the perihelion precession, $\dot{\omega}_{\ppK}$ (related to $k_{\ppK}$), the gravitational redshift due to the pulsar's companion, $\gamma_{\ppK}$ , the range and shape of the Shapiro time delay, $r_{\ppK}$ and $s_{\ppK}$, and the rate of decay of the orbital period, $\dot{P_b}^\ppK$. We will restrict ourselves in this paper to considering these five parameters. 

Our ability to test GR using these ppK parameters depends on the precision with which we can measure them. The current uncertainties in the measured values of these parameters are listed in Table \ref{table:ppKparam}.

\begin{table}[ht]
\begin{center}
    \begin{tabular}{c|c|c}
    \hline\hline
  Parameter                & Effect                          & Measured value \\ \hline 
$\dot{P}_b^\ppK$                & orbital decay                 & $-1.252(17)\times10^{-12} $   \\ 
 $r_{\ppK}$                 & range of Shapiro delay  & $6.21(33)(\mu s)$          \\ 
$s_{\ppK}$                 & shape of Shapiro delay  & $0.99974(+16,-39)$   \\ 
$\dot{\omega}_{\ppK}$ & periastron precession    &$016.89947(68)(^{\circ}$/yr)  \\
$\gamma_\ppK$           & gravitational red-shift   & $0.3856(26)$ (ms) \\ \hline\hline
  \end{tabular}
   
    \caption{Uncertainty in measured values for PSR J0737-3039A. \cite{Kramer:2006nb} \label{table:ppKparam}}
    
\end{center}
\end{table}

\subsection{Parameterized post-Einsteinian \label{sec:ppE}}
In the near future, the detection of GWs will open a new window for testing GR. There are essentially two approaches for doing so. One method, a \emph{top-down} approach, demands that we have a particular alternative gravity theory that we wish to test. In this method, we could calculate what GWs would look like in this theory, and develop non-GR GW templates for data analysis. For a given signal, one could then calculate the Bayesian odds ratio\footnote{The Bayesian odds ratio, or \emph{Bayes factor}, between two models, A and B, is the betting odds that model A is supported by the data better than model B. A Bayes factor of $3$ in favor of model A would indicate that the data shows a $3$ to $1$ preference for model A.} between this specific theory and GR,  and in this way decide which theory is better supported by the data. The advantage to this method is that we would have the full equations of motion for the model, and be able to answer theoretical questions such as well-posedness, in addition to being able to predict many observables. On the other hand, there is no particularly compelling alternative to GR presently known, and the effort involved in fleshing out all possible contenders is highly non-negligible.

In contrast, the second method for testing GR is a \emph{bottom-up} \cite{Yunes:2013dva} approach . In this approach, one uses experimental data to learn about a possible gravitational theory. If there is an indication of deviation from GR, these data then motivate the development of an alternative to GR. In order to use GWs for this type of analysis, we need a set of waveform templates that do not assume that GR is the correct theory of gravity.

With these templates as their aim, Yunes and Pretorius \cite{Yunes:2009ke} developed the ppE family of waveform templates. They focused on creating templates for the Fourier transform of the quadrupole GW strain signal from a system of two inspiraling, non-spinning, compact objects in quasi-circular orbits. In the future, the restriction to circular, non spinning systems can be relaxed, building on work done in Ref.~\cite{Yunes:2009yz} for eccentric systems in GR and Ref.~\cite{Chatziioannou:2013dza} for spinning systems in GR. The restriction to the quadruple mode has already been lifted by Chatziioannou et al in \cite{Chatziioannou:2012rf}. 

The full waveform from two coalescing bodies is typically split into three phases - inspiral, merger, and ringdown. ppE templates have been developed for all three phases, but here we restrict ourselves to the inspiral only. The inspiral is the part of the waveform that is generated while the two bodies are still widely separated, and thus slowly spiraling towards each other due to the emission of GWs. The definition of the end of inspiral is somewhat arbitrary, but we follow typical convention and define the transition from inspiral to merger as occurring at the innermost stable circular orbit of the system in center of mass (COM) coordinates. The simplest, quadrupole ppE inspiral templates have the form:
\be
\tilde{h} (f) = \tilde{h}^{GR}\cdot (1+\alpha_{\ppE}  u^{a}) e^{i\beta_{\ppE} u^{b}}, \quad\quad u = (\pi \mathcal{M} f)^{1/3},
\label{eq:ppEtemp}
\ee
where $\tilde{h}^{GR}$ is the GW waveform in GR, $\mathcal{M}$ is the chirpmass of the system, $\mathcal{M} = (m_1 m_2)^{3/5}/(m_1+m_2)^{1/5}$, and $f$ is the GW frequency. These simple ppE modified waveforms consist of an additional amplitude term, $\alpha_{\ppE} u^{a}$, and an additional phase term, $\beta_{\ppE}  u^{b}$, relative to GR. We refer to $\alpha_{\ppE} $ and $\beta_{\ppE} $ as the \emph{strength} parameters of the ppE deviations.

The ppE templates are constructed by introducing parameterized modifications to both the binding energy and the energy balance equations of GR. Both types of modifications lead to changes in the GW phase, which leads to a degeneracy if we are only sensitive to the phase. This means that if a deviation from GR is detected in the phase, it is impossible to determine from only this measurement whether the deviation is from the conservative or dissipative sector. The combination of GW measurements with other experiments, as well as more sensitive measurements of GW amplitude, could possibly lift this degeneracy.

The ppE waveforms cover all known inspiral waveforms from specific alternative theories of gravity \cite{Cornish:2011ys} that are analytic in the frequency evolution of the GWs. Some specific examples are listed in Table \ref{Table:ppEpar}.
\begin{table}
\begin{center}
    \begin{tabular}{c|c|c|c|c}
    \hline\hline
  Theory               & $a$       & $\alpha_{\ppE} $ & $b$ & $\beta_{\ppE} $ \\ \hline 
 Variable G(t)     &$-8 $        & $\alpha_{\ppE}' $          & $-13$& $\beta_{\ppE}'$\\ 
  Brans-Dicke      &$-2 $        & $\alpha_{\ppE}' $        & $-7$& $\beta_{\ppE}' $\\ 
 Dynamical Chern-Simons     & a         & $0$          & $-1$& $\beta_{\ppE}' $\\ \hline\hline
  \end{tabular}
    \caption{The values that the ppE parameters take on in various non-GR theories \cite{Yunes:2013dva}. \label{Table:ppEpar}}

\end{center}
\end{table}

\section{ppN-ppE correspondence \label{sec:ppNppE}}

The first correspondence we calculate is that between the ppE and ppN parameters. In particular, we are interested in how the ppE phase parameters can be related to the ppN metric. The ppE amplitude parameters can also be related to ppN corrections to the metric. These corrections, like the phase ones, enter at first PN order. Because we are much more sensitive to the GW phase in analyzing GW data, we typically do not include any PN corrections to the amplitude, and so we do not show this calculation. 

To calculate the GW phase, we use 
\be
\phi(t) = 2\pi \int_{t_0}^t dt' \int_{t_0'}^{t'} \dot{f}(t'') dt''.
\ee
We can calculate $\dot{f}(t)$ from 
\be
\frac{df}{dt} = \frac{df}{dE}\frac{dE}{dt},
\label{Eq:fdot}
\ee
where E is the binding energy of the binary, and $\dot{E}$ is the GW luminosity. This is the same technique used by Yunes and Pretorius in developing the ppE framework \cite{Yunes:2009ke}. We find the integrand in Eq.~(\ref{Eq:fdot}) by calculating the binding energy as a function of velocity, using the ppN modified Kepler's law to change this to a function of frequency, and then inverting this expression so we have the frequency as a function of energy. We then use the standard GR expression for the GW luminosity, coupled with our non-GR expression for $df/dE$, and we can calculate the phase of the gravitational waveform. 

Each of these steps needs to be carried out to consistent PN order. In our case, we are only interested in the 1PN correction to GR, as this is the order to which the ppN framework is valid. In the COM frame of the two-body system, the binding energy in the ppN formalism, correct to 1PN order, is \cite{Will:1993ns}
\begin{multline}
 E_{COM} = -\frac{M \mu}{r_{12}} + \frac{1}{2}\mu v^2 + \frac{3}{8}\frac{\mu}{M}v^4(M-3\mu) 
 +\frac{M\mu}{2r_{12}}\\ \times\Big[ (2\gamma_{\ppN}+1)v^2  + \frac{M}{r_{12}}(2\beta_{\ppN} -1) \\ + v^2 \frac{\mu}{M}\left(1+\alpha_1^{\ppN}-\alpha_2^{\ppN}\right)\Big].
 \label{eq:ECOM}
\end{multline}
Here $M$ is the total mass, $\mu$ is the reduced mass, $r_{12}$ is the separation distance between the two bodies, and $v$ is the magnitude of the relative velocity between the two bodies in the COM frame.

Next, and again from \cite{Will:1993ns}, the magnitude of the acceleration between the two bodies, in the COM frame, in the ppN formalism, correct to 1PN order, is
\begin{multline}
a = r_{12}\omega^2 -\frac{M}{r_{12}^2}\bigg\{1 - \frac{M}{r_{12}} \bigg[2\beta_{\ppN} + \gamma_{\ppN} \\- \eta\bigg(1+ \frac{\alpha_1^{\ppN}-\alpha_2^{\ppN}}{2}\bigg)\bigg] \bigg\}.
\label{eq:aCOM}
\end{multline}
 Note that the ppN parameter, $\xi$, having to do with preferred location effects, does not appear in either the binding energy or the acceleration. Due to symmetry, the system being considered would need to consist of at least three bodies for these effects to be non-zero. This means that the set of ppN parameters relevant to our problem has been reduced from 10 to 4, which are $\beta_{\ppN}, \gamma_{\ppN}, \alpha_1^{\ppN}$, and $\alpha_2^{\ppN}$.

 To proceed, we must re-write Eq. (\ref{eq:ECOM}) as a function of the GW frequency, $f$. To do so, we use the fact that, in a circular orbit, correct to 2.5 PN order, $v = r_{12} \omega$ and $a = r_{12}\omega^2$. We can thus us Eq. (\ref{eq:aCOM}) to find a relation between $v$ and $r_{12}$, which we can then use to re-write Eq. (\ref{eq:ECOM}). We then have an expression for the binding energy in terms of GW frequency:
\begin{multline}
E = -\frac{1}{2}\mu (2\pi M f)^{2/3} \bigg(1+ (2\pi M f)^{2/3}\\ \times  \bigg\{-\frac{3}{4} + \frac{2}{3}(\beta_{\ppN}-\gamma_{\ppN})  - \eta\bigg[\frac{1}{12}+ \frac{5}{3}(\alpha_1^{\ppN}-\alpha_2^{\ppN})\bigg]\bigg\} \bigg).
\label{eq:E}
\end{multline}
We then differentiate Eq. (\ref{eq:E}) with respect to $f$, and invert the result to find $df/dE$
\begin{multline}
\frac{df}{dE} = -\frac{\mu}{3}(2\pi M)\bigg( (2\pi M f)^{-1/3} + 2 (2\pi M f)^{1/3} \\  \times \bigg\{ -\frac{3}{4} + \frac{2}{3}(\beta_{\ppN} - \gamma_{\ppN}) - \eta \bigg[\frac{1}{12}+\frac{5}{3}(\alpha_1^{\ppN}-\alpha_2^{\ppN}) \bigg]\bigg\}\bigg).
\label{eq:dfdE}
\end{multline}
where we have expanded in $M f \ll 1$.

Lastly, we need the energy carried away from the system as GW luminosity. At this point, we make the assumption that the luminosity in GWs for alternative theories can be calculated from the mass and current multipoles of the system in the same way as in GR. Put another way, we are relating the changes in the conservative, as opposed to dissipative, sector of gravity to effects seen in the waveforms. This is one instance in which it is clear that the ppE-ppN mapping is not perfect, because the ppE formalism includes changes to the dissipative sector as well as to the binding energy, whereas the ppN formalism is concerned only with the conservative sector. 

In GR, and to 1PN order, the GW luminosity is given by~\cite{lrr-2006-4}
\be
L_{GW} = \frac{1}{5}M_{ij}^{(3)}M_{ij}^{(3)} + \frac{1}{189}M_{ijk}^{(4)}M_{ijk}^{(4)} + \frac{16}{45}S_{ij}^{(3)}S_{ij}^{(3)},
\label{eq:lummult}
\ee
where the mass quadrupole moment is $M_{ij}$, the mass octupole is $M_{ijk}$, and the current quadrupole is $S_{ij}$; the superscript $(n)$ implies $n$ time derivatives; and the source multipole moments of the binary are calculated in the standard way \cite{lrr-2006-4}. 
\begin{table*}[ht]
\begin{center}
    \begin{tabular}{c|c|c|c|c|c}
    \hline\hline
    Theory       & Un-known functions or constants&$\bar{\gamma}_\ppN$ & $\bar{\beta}_\ppN$ & $\alpha_1^\ppN$ & $\alpha_2^\ppN$\\ \hline
  Brans-Dicke & $\omega_{BD}$                             & $\frac{1}{\omega_{BD}}$ & 0                          & 0                    & 0 \\ 
  General scalar-tensor & $A(\varphi),V(\varphi)$      & $\frac{1}{\omega} $        & $\frac{1}{4 \omega^3}\frac{d\omega}{d\phi}\Big|_{\phi_0}$               & 0                    & 0 \\ 
  Einstein-Aether         & $c_1,c_2,c_3,c_4$            &0					 & 0                          & $\alpha_1'$          & $\alpha_2'$ \\ \hline\hline
           \end{tabular}
    \caption{Values of ppN parameters for a selection of alternative theories. These expressions are in the large $\omega$ limit. \cite{lrr-2006-3,PhysRev.124.925,Jacobson:2008aj,Faraoni:1998qx}   \label{table:ppNparalt}}
\end{center}
\end{table*}
Using these assumptions, we arrive at an expression for the GW luminosity,
\be
\frac{dE}{dt} = \frac{32}{5} \eta^2 M^2 r^4 (2\pi f)^6\left[ 1 + \frac{M}{r_{12}}\left(-\frac{10}{7} + \frac{16}{7} \eta \right) \right].
\label{eq:Lum}
\ee
We then use the mapping between $v$ and $r_{12}$ to re-write this equation as:
\begin{multline}
\frac{dE}{dt} = \frac{32}{5} \eta^2 (\pi M f)^{10/3} \bigg(1 + (\pi M f)^{2/3} \\ \times \Big\{\frac{97}{336} - \frac{8}{3}\beta_{\ppN} - \frac{4}{3}\gamma_{\ppN} + \eta \big[-\frac{35}{12} + \frac{2}{3}(\alpha_1^{\ppN}-\alpha_2^{\ppN})\big] \Big\} \bigg).
\label{eq:lum2}
 \end{multline}
Multiplying Eq. (\ref{eq:lum2}) and Eq. (\ref{eq:dfdE}), and again expanding in $Mf\ll1$, we arrive at
\begin{multline}
\frac{df}{dE}\frac{dE}{dt} = \frac{96}{5} \pi^{8/3}f^{11/3}\mathcal{M}^{5/3} \bigg(1 +(\pi M f)^{2/3}\\ \times \bigg\{\frac{601}{336} -4\beta_{\ppN} - \eta \bigg[\frac{11}{4} - 4(\alpha_1^{\ppN} - \alpha_2^{\ppN})\bigg]\bigg\} \bigg).
\label{Eq:firstInt}
\end{multline}
We next integrate Eq. (\ref{Eq:firstInt}) once, and then invert perturbatively to find
\begin{multline}
t(f)  = t_c-5 (8\pi f)^{-8/3} \mathcal{M}^{-5/3}  \bigg\{1 + (\pi M f)^{2/3}\\ \times \bigg( -\frac{601}{252} + 
\frac{16}{3} \beta_{\ppN} + \eta \bigg[\frac{11}{3}-\frac{16}{3}(\alpha_1^{\ppN}-\alpha_2^{\ppN}) \bigg]\bigg)\bigg\}.
\label{eq:toff}
\end{multline}
Here $t_c$ is the time of coalescence.  We integrate Eq. (\ref{eq:toff}) again to find $\phi(t)$, and then replace $t$ with Eq.~(\ref{eq:toff}) to arrive at
\begin{multline}
\phi(f) = \phi_c - 2(8\pi \mathcal{M} f)^{-5/3}\bigg(1+(\pi M f)^{2/3}\\\times \bigg\{-\frac{3005}{1008}+\frac{20}{3}\beta_{\ppN}+ \eta \Big[\frac{110}{24} - \frac{20}{3}(\alpha_1^{\ppN}-\alpha_2^{\ppN}) \Big]\bigg\} \bigg),
\end{multline}
where $\phi_c$ is the phase at coalescence.

Finally, we can use the stationary phase approximation (SPA) to find the phase of the GW in the Fourier domain. To calculate this, we use $\Psi_{\SPA} = 2\pi f t(f) - \phi(f) -\pi/4 $, which gives
\begin{multline}
\Psi_{\SPA}(f) = 2 \pi f t_c - \phi_c - \frac{\pi}{4}\\ + \frac{3}{128}u^{-5}\bigg\{1 + u^{2} \bigg[\frac{20}{9 \eta^{2/5}}\bigg(\frac{743}{336} +\frac{11}{4}\eta\bigg)\bigg] \bigg\} \\+u^{-3} \frac{5}{24 \eta^{2/5}} \big[\bar{\beta}_{\ppN} -\eta (\alpha_1^{\ppN}-\alpha_2^{\ppN})\big].
\label{eq:SPA}
\end{multline}
where recall that $u = (\pi \mathcal{M} f)^{1/3}$.

The last term in  Eq. (\ref{eq:SPA}) represents a ppE correction with $b=-3$, and strength parameter 
\be
\beta_{\ppE}^{b=-3} = \frac{5}{24 \eta^{2/5}}\big\{\bar{\beta}_{\ppN} -\eta (\alpha_1^{\ppN}-\alpha_2^{\ppN})\big\}.
\label{eq:ppEconst}
\ee
Note that in the test-particle limit, when $\eta \rightarrow 0$, the $\beta_{\ppE}$ and $\bar{\beta}_{\ppN}$ parameters are simple rescalings of each other. It is also interesting to note that the parameter $\gamma_{\ppN}$ does not appear in the final expression for the phase. This tells us that within alternative theories of gravity that do not alter the value of $\beta_{\ppN}, \alpha_1^\ppN$, or $\alpha_2^\ppN$, there will be no phase corrections at first PN order.

Table \ref{table:ppNparalt} lists the values of the ppN parameters for a selection of alternative theories of gravity. None of these theories have $\bar{\beta}_\ppN$ that is strongly different from the GR value of $\bar{\beta}_\ppN = 0$. This tells us that if a significant departure from GR is detected at 1 PN order, it must come from the dissipative sector of these alternative theories, or from a different theory altogether.

\section{ppN-ppK Correspondence \label{sec:ppNppK}}
Now we derive the connections between the ppN metric parameters and the ppK timing parameters. The arrival time of a pulse of EM radiation at the Earth is expressed schematically in Eq. (\ref{eq:timingschem}). The Roemer time delay for a binary pulsar, as stated above, includes not only the (non-GR) modulation due to the motion of the Earth around the Sun, but also modulations due to the orbital motion of the binary itself. This orbital motion includes GR corrections at 1PN order, which lead to perihelion precession, $\dot{\omega}$, one of the ppK parameters. This parameter can be related to the ppN parameters by calculating the equations of motion for the pulsar using the ppN metric. This calculation is done by Will \cite{Will:1993ns}, and, in the semi-conservative theories of gravity that we are considering, is equal to 
\begin{multline}
<\dot{\omega}> = \frac{(2M)^{2/3}\pi^{5/3}}{(e^2 -1)P_b^{5/3}}\big[ 2 (1+\bar{\beta}_{\ppN} -2\bar{\gamma}_{\ppN}) \\- \eta(2\alpha_1^{\ppN}-\alpha_2^{\ppN})\big].
\end{multline}
where the angled brackets indicate integration over one orbit.

In order to be consistent with our other analyses, this result, and all results in this section, neglects any contributions from the self-gravity or structure of the pulsars. These types of effects have been explored in \cite{Wex:2007ct}.

The Shapiro time delay is calculated from the formula for null geodesics. In GR, to the appropriate order, this is simply
\be
dt = \big[1- 2\phi(x)\big] dx,
\ee
where $\phi(x)$ is the Newtonian gravitational potential due to the pulsar's companion, $m_c/r_{12}$, with $m_c$ the mass of the companion, and $r_{12}$ the separation between them\cite{Maggiore:1080850}. In the ppN formalism, this equation becomes
\be
dt = \Big[1-(1+\gamma_{\ppN})\phi(x)\Big]dx.
\ee 
Integrating this over the path of the photon from the pulsar, past its companion, and to the solar system barycenter, we get
\begin{multline}
\Delta_S = - 2 r_{\ppK} \log \Big\{ [1- e \cos(u)] - s \big[\sin (\omega) (\cos (u) -e)\\ + \sqrt{1-e^2}\cos (\omega) \sin (u)\big] \Big \},
\end{multline}
which is the same as in GR, except now
\be
r_\ppK = m_c\left(1+\frac{\bar{\gamma}_{\ppN}}{2}\right),
\ee
is the range of the Shapiro delay. The shape, $s$, is still equal to $\sin(\iota)$, as in GR, but it is generally written in terms of the two masses in the binary, $m_p$ and $m_c$, using Kepler's law. Kepler's law is altered at 1PN order from the GR expression, but this alteration should not be considered, as $r_{\ppK}$ is already a 1PN correction. Thus, the formula for $s$ in terms of the masses of the system is unaltered from GR:
\be
s = \frac{x}{(m_c+m_p)^{1/3}}\left(\frac{P_b}{2\pi}\right)^{-2/3},
\ee
where recall that $x$ is the projected semi-major axis.

Finally, we calculate the Einstein time delay, which is really just the gravitational redshift. This redshift arises from the time dilation experienced by a photon as it travels out of a gravitational potential well. It is expressed as the rate of change of proper time with respect to coordinate time
\begin{table*}[ht]
\begin{center}
\begin{tabular}{c|c}
\hline \hline
ppK Parameter & ppN expression \\ \hline
$\gamma_{\ppK}$ &$\frac{P_b}{2\pi}\frac{e}{2 a(m_c+m_p)}\big\{3 m_c^2 +m_cm_p+(\bar{\gamma}_{\ppN}+1)(m_c^2 +m_cm_p)\big\}$ \\ 
$r_{\ppK}$  & $\frac{1}{2} (\bar{\gamma}_{\ppN}+2)m_c$ \\ 
$<\dot{\omega}>$ & $\frac{(2M)^{2/3}\pi^{5/3}}{(e^2 -1)P_b^{5/3}}\left\{ 2 (1+\bar{\beta}_{\ppN} -2\bar{\gamma}_{\ppN}) - \eta(2\alpha_1^{\ppN}-\alpha_2^{\ppN})\right\}$\\ \hline\hline
\end{tabular} 

\caption{ppK parameters expressed as combinations of the ppN parameters.\label{Table:ppNppK}}
\end{center}
\end{table*}
\be
d\tau^2 = \left[ 1 + (1+\gamma_{\ppN})\phi(x)\right] dt^2 - \left[1-(1+\gamma_{\ppN})\phi(x) \right]dx^2,
\ee
which leads, in the weak-field approximation, to
\be
\frac{d\tau}{dt} = 1 + \frac{1+\gamma_{\ppN}}{2}\phi(x) - \frac{1}{2}v_p^2,
\label{eq:dtaudt}
\ee
where $v_p = dx/dt$ is the velocity of the pulsar. We can next use the virial theorem to replace $v_p^2$. 
\be
\frac{1}{2}v_{12}^2 = \frac{m_p+m_c}{r_{12}} - \frac{m_p+m_c}{2a}, \quad v_p = \frac{m_c}{m_p+m_c}v_{12},
\label{eq:virial}
\ee
with the separation for eccentric orbits given by $r_{12} = a(1-e \cos u)$, and $v_{12}$ the relative velocity between the two bodies. Just as with Kepler's law, there are 1PN corrections, but, in order to keep all terms to the proper order, we use the Newtonian approximation. We use the relationships in Eq. (\ref{eq:virial}) to replace $v_p$ in Eq. (\ref{eq:dtaudt}) with an expression in terms of $v_{12}$.

Next, we want to change variables from the eccentric anomaly, $u$, to $P_b$. Differentiating Kepler's equation, Eq. (\ref{eq:Kepler}), and neglecting any terms that go as $\dot{P}_b$, we arrive at
\be
\frac{du}{dt} = \frac{2\pi}{P_b}\frac{1}{1- e \cos u}.
\label{eq:transf}
\ee
We change coordinates using Eq. (\ref{eq:transf}) and the fact that $d\tau/dt = (d\tau/du)(du/dt)$, and, finally, Eq. (\ref{eq:dtaudt}) becomes
\begin{multline}
\frac{2\pi}{P_b}\frac{d\tau}{dt} = \left\{1 - \frac{m_c[2m_c+m_p+\gamma_{\ppN}(m_c+m_p)]}{2a(m_c+m_p)} \right\} \\ \times\left\{1- e \cos u \left[1+ \frac{3m_c^2 + m_cm_p + \gamma_{\ppN}(m_c^2 + m_cm_p)}{2a(m_c+m_p)} \right] \right\}.
\end{multline}
The part of this expression that is unmodulated with $u$ is not detectable. We can absorb it into a rescaling of the proper time
\be
\tau \rightarrow \tau\times\left\{ 1 - \frac{m_c[2m_c + m_p + \gamma_{\ppN}(m_c+m_p)]}{2a(m_c+m_p)}\right\}.
\ee
This leaves us with the formula for the gravitational redshift in the standard form
\be
\frac{d\tau}{dt} = \frac{P_b}{2\pi}(1 - e \cos u ) - \gamma_{\ppK}  \cos u,
\ee
where we find that $\gamma_\ppK$ is related to ppN parameters via
\be
 \gamma_{\ppK} = \frac{P_b}{2\pi}e\frac{3 m_c^2 +m_cm_p+\gamma_{\ppN}(m_c^2 +m_cm_p)}{2 a(m_c+m_p)}.
\ee
In summary, we have a correspondence between the ppN parameters and four of the ppK parameters, in semi-conservative theories of gravity in a reference frame at rest with respect to any universal reference frame, and neglecting effects due to self-gravity and structure of the pulsars. This correspondence is summarized in Table \ref{Table:ppNppK}. Because of the combinations of ppN parameters that appear in these expressions, it is not possible to use the results from Sec. \ref{sec:ppNppE} and re-write them entirely in terms of ppE parameters.

\section{ppE-ppK Correspondence \label{sec:ppKppE}}

The final piece missing from our correspondence puzzle is the relationship between the decay of the orbital period of a binary system, the ppK parameter $\dot{P_b}$, and either ppN or ppE parameters. The correspondence with ppE parameters was tackled by Yunes and Hughes  \cite{2010PhRvD82h2002Y}. We here carry out a similar calculation.

Assuming that the binding energy of a system is the same as in GR, the decay rate of a binary system can be calculated via
\be
\frac{\dot{P_b}}{P_b} = -\frac{3}{2} \frac{\dot{E}}{E_b},
\label{eq:decay}
\ee
where $\dot{E}$ is the energy carried away by GWs, and $E_b$ is the binding energy of the system. 

Yunes and Hughes assumed in their calculation that only the dissipative sector is affected by the ppE parameters. This is because the functional form of the ppE phase corrections does not depend on whether the binding energy, the GW luminosity, or both are modified from GR. This degeneracy makes it impossible to determine whether the ppE changes in the GW phase arise from the dissipative or conservative sector. Thus, we first take the presence of ppE parameters in the phase of the GW to come only from the expression for $\dot{E}$ \cite{2010PhRvD82h2002Y}:
\be
\dot{E}  = \dot{E}_{\GR}\left( 1 + \pi^2 \mathcal{M}^2 u^{-1}\frac{d^2 \Psi_{\GR}}{df^2} \right),
\ee
where $\Psi_{\GR}$ is the phase of the GW in GR.

The ppE corrections have only been calculated assuming circular binaries. However, all known pulsars are in eccentric orbits. The corrections to $\dot{E}$ from eccentricity are known in GR
and have the form
\be
\dot{E}_{\GR} = -\frac{32}{5}\eta^2 \frac{M^5}{a^5}(1-e^2)^{-7/2}\left(1+\frac{73}{24}e^2 + \frac{37}{96}e^4\right).
\ee
The same type of corrections, however are not known in the ppE framework, and will necessarily involve the introduction of new ppE parameters. Fortunately, the eccentricity of some pulsar systems is small enough, that we can accurately model the ppE parameters as expansions in small $e$, where we keep only the lowest order term in our calculations. 

With this assumption, we can use Eq. (\ref{eq:decay}) to show that the expression for the orbital decay rate becomes \cite{2010PhRvD82h2002Y}
\be
\left(\frac{\dot{P}_b}{P_b}\right)_{Phase} = \left( \frac{\dot{P}_b}{P_b} \right)_{\GR} \left[1+\frac{48}{40}\beta_{\ppE}  b(b-1)u^{b+5}\right],
\ee
when corrected with phase ppE parameters. The term $(\dot{P}/P)_{\GR}$ stands for the orbital decay rate in GR for an eccentric inspiral. Because the observed value of $\dot{P}/P$ is very close to the GR prediction, we can write $(\dot{P}_b/P_b)_{obs} = (\dot{P}_b/P_b)_{\GR}(1+\delta)$, where $\delta$, the observational error, is small. 

Because the periods of binary pulsars have been observed to decay at the GR rate, the ppE strength parameter must satisfy \cite{2010PhRvD82h2002Y}
\be
|\beta_{\ppE}| \le \frac{40}{48 |b||b-1|}\frac{\delta}{u^{b+5}} \,\,.
\label{eq:ppKbeta}
\ee

As stated, the preceding calculation was done with the assumption that only the GW luminosity of the pulsar system was different from GR. The binding energy was assumed to be the same as in GR. This is the opposite of what we assumed in Sec. \ref{sec:ppNppE}, in which only the binding energy was altered. We now calculate the relation between $\beta_\ppE$ and $\dot{P}_b$, but this time assuming that it is the conservative sector that is altered from GR.

Following \cite{Chatziioannou:2012rf}, we parameterize the binding energy of a binary system as
\be
E = E_\GR \bigg[1+ A \bigg(\frac{M}{r_{12}} \bigg)^q \bigg],
\ee
where $A$ is small, and therefore $E$ differs from $E_\GR$ by only a small perturbation. This binding energy leads to a modified Kepler's law:
\be
\omega^2 = \frac{M}{r_{12}^3}\bigg[ 1 + \frac{1}{2} A q \bigg( \frac{m}{r}\bigg)^q \bigg],
\label{eq:KepppEppK}
\ee
which we can use to re-write the energy in terms of the orbital period of the pulsar system:
\begin{align}
E =& -\frac{1}{2} \eta^{-2/5}\bigg(\frac{2 \pi \cal{M}}{P_b} \bigg)^{2/3}\nonumber\\ & \times \bigg[1 - \frac{1}{3} A (5q - 6) \eta^{-2q/5} \bigg(\frac{2\pi \cal{M}}{P_b} \bigg)^{2q/3} \bigg].
\label{eq:EnofP}
\end{align}

We can find an expression for $\dot{P}_b/P_b$ by differentiating Eq. (\ref{eq:EnofP}) with respect to time. This will give us an expression in terms of the GW luminosity, $\dot{E}$. Although we are not explicitly changing the GR expression for this luminosity, $\dot{E}$ is modified when we use Eq. (\ref{eq:KepppEppK}) to relate $r_{12}$ to $P_b$.
\begin{align}
\dot{E} =& \frac{32}{5} \bigg(\frac{2 \pi \cal{M}}{P_b} \bigg)^{10/3}\nonumber\\&\times\bigg[1 - \frac{1}{3} A q \eta^{-2q/5} \bigg(\frac{2\pi\cal{M}}{P_b}\bigg)^{2q/3} \bigg].
\label{eq:Lum3}
\end{align}

Differentiate Eq. (\ref{eq:EnofP}) with respect to time, and replace $\dot{E}$ with Eq. (\ref{eq:Lum3}) and arrive at
\be
\frac{\dot{P}_b}{P_b} = \bigg(\frac{\dot{P}_b}{P_b} \bigg)_\GR \bigg[1- \frac{1}{3} A \eta^{-2q/5}(5q^2 - 2q -6)u^{2q} \bigg],
\ee
where we have expanded in $A\ll1$.

The final step is to relate $q$ and $A$ to the more standard ppE parameters, $b$ and $\beta_\ppE$. The relations are \cite{Chatziioannou:2012rf}
\be
2q = b+5,
\ee
\be
-A (5q^2 - 2q -6)\eta^{-2q/5} = \frac{32}{5} \beta_\ppE (4-q)(5-2q).
\ee

With these replacements, we finally arrive at
\be
\frac{\dot{P}_b}{P_b} = \bigg(\frac{\dot{P}_b}{P_b} \bigg)_\GR \bigg[1 + \frac{16}{15}\beta_\ppE b (b-3)u^{b+5} \bigg],
\ee
which leads to the constraint on $\beta_\ppE$:
\be
|\beta_\ppE|\le \frac{15}{16} \frac{\delta}{|b||b-3|u^{b+5}}.
\label{eq:ppKbeta2}
\ee

Both Eq. (\ref{eq:ppKbeta}) and Eq. (\ref{eq:ppKbeta2}) are constraints on the ppE strength parameter associated with the phase of a GW. In the first case, this constraint comes from the assumption that the GW luminosity of the system is not as described by GR, while in the second case it is the binding energy that is changed from the GR expression. The fact that both approaches lead to constraints on the GW phase illustrates a degeneracy in the ppE formalism. Both changes to the binding energy and changes to the luminosity of a system lead to the same type of non-GR terms in the GW phase. It is impossible to tell if a ppE term in a GW signal arises from the conservative or dissipative sector of a gravitational theory, or some combination of the two, if one is sensitive only to the phase.

\section{Current Constraints \label{sec:Constraints}}
With expressions for the correspondences between parameters in the different systems in hand, we can calculate current constraints on the ppE parameters from known constraints on ppN and ppK parameters. Sec. \ref{sec:ppNppE}, Eq. (\ref{eq:ppEconst}) gives the bound on $\beta_\ppE$ from the constraints on ppN parameters. Using the best constraints on the ppN parameters from Table \ref{Table:ppNlimits}, in Fig. \ref{ppElimppN} we plot this limit as a function of mass ratio. We see that $\beta_\ppE < 0.008$ from Solar System tests.

\begin{figure}[ht]
\begin{center}
\begin{tabular}{cc}
\hspace*{-8mm} \includegraphics[width =9.5 cm]{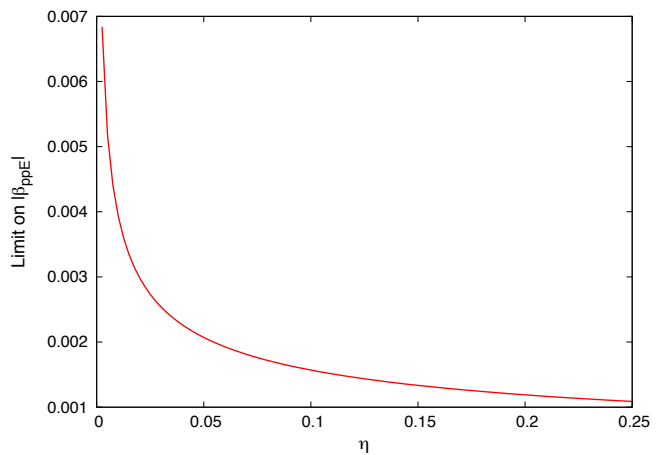} 
\end{tabular}
\end{center}
\vspace*{-0.4in}
\caption{\label{ppElimppN} Limits that can be placed on the ppE strength parameter, $\beta_{\ppE}$, using the known limits on ppN parameters, and the results from Sec. \ref{sec:ppNppE}. Regions above the curve are ruled out. The limits are a function of mass ratio, and the ppE $b$ parameter is set to $b=-3$ in this plot.}
\end{figure}
Similar bounds on $\beta_{\ppE}$ from GW detections (with signal-to-noise ratio of $20$) were calculated in \cite{Cornish:2011ys}. For both a $2:1$ and a $3:1$ mass ratio, the authors found a limit of $\beta_{\ppE} \lesssim 0.003$, which is comparable to current bounds from the Solar System.However, with multiple detections, the GW bounds should eventually surpass the latter.

\begin{table*}[ht]
\begin{center}
    \begin{tabular}{c|c }
    \hline\hline
    Source of constraint & Constraint \\ \hline
    Pulsar - $\dot{E}$ non-GR & $\beta_\ppE \le 215$ \\
    Pulsar - $E_b$ non-GR & $\beta_\ppE \le 182$  \\
    Solar-System tests & $\beta_\ppE  \lesssim0.001$ \\
    Anticipated GW detections &  $\beta_\ppE  \lesssim 0.008$ \\  \hline\hline
           \end{tabular}
   
    \caption{ \label{Table:Constraints} The current constraints that can be placed on $\beta_\ppE$ from Solar System and binary pulsar tests. These values are only for $b = -3$.}
\end{center}
\end{table*}

The observed values for the ppK parameters from PSR J0737-3039 can similarly be used to place constraints on the ppE parameters, this time using results from Sec. \ref{sec:ppKppE}. For this pulsar system, the uncertainty in the measurement of $\dot{P}_b/P_b$ is $\delta = 0.017 \times 10^{-12}/(1.242\times10^{-12}) \backsimeq10^{-2}$. The chirp mass is $\mathcal{M} \backsimeq 5.5399\times10^{-6}$ s, and the GW frequency is $f\backsimeq2.263842976\times10^{-4}$ Hz. Using these values in Eq.~(\ref{eq:ppKbeta}) and Eq.~(\ref{eq:ppKbeta2}), we find that a GW measurement can set better constraints than the pulsar measurements starting around $\sim b = -4$. This means that there are regimes in which measurements of ppK parameters can help to constrain ppE parameters, and also regimes in which the opposite is true. This is consistent with the conclusions of \cite{Cornish:2011ys}.

The current constraints for $\beta_\ppE$, with $b=-3$, from Solar System and pulsar experiments are listed in Table \ref{Table:Constraints}, as well as the anticipated constraint from future GW detections. The bounds from pulsar data do not depend strongly on whether it is the conservative or dissipative sector that is changed from GR. Both of these constraints are weaker than the Solar System and GW bounds, which differ from each other by a factor of $\sim3$.

\section{Conclusion \label{sec:Conc}}

The many parameterizations of gravitational theory that have been developed over the years are designed for, and thus ideally suited to, testing the nature of gravity in quite different situations. We know, though, that the underlying theory of gravity that describes the universe we live in is the same in our solar system as it is in binary pulsar systems and colliding black holes. We should therefore be able to learn about the ppE parameters from our knowledge of ppN and ppK parameters, and vice versa. 

In this paper, we have found correspondences between the ppE, ppN, and ppK parameters that allows us to apply constraints from one formalism to the parameters in the others. In addition to finding the connections between the parameters in the different formalisms, in this work we have found that alternative theories of gravity that do not alter the $\beta_{\ppN}, \alpha_1^\ppN,$ or $\alpha_2^\ppN$ parameters do not result in 1PN corrections to the GW phase. We also found that the bounds we will be able to place on deviations from GR at the 1PN level using GWs will be comparable to those already known from solar system tests.

The correspondences that we have calculated are not perfect. The ppN-ppE correspondence assumes semi-conservative theories of gravity, in a reference frame at rest with respect to any universal preferred frames. The ppN-ppK correspondence makes the same assumptions. And the ppE-ppK correspondence is only perfectly accurate for circular binaries. Finally, both the ppN-ppE and ppE-ppK correspondences assume that the generation of GWs, used to calculate the luminosity, is the same as in GR.

Future work could focus on relaxing some of the assumptions we used in this analysis. For instance, by allowing for changes in Eq. (\ref{eq:lummult}) by including source or current multipoles that are not present in GR. It may also be possible to introduce eccentricity into the ppE formalism, which could improve the accuracy of the ppE-ppK correspondence.

\acknowledgments
We acknowledge Paul Baker and Katerina Chatziioannou for useful conversations. L.~S. and N.~C. were supported by NSF grant PHY-1205993 and NASA grant NNX10AH15G. N. Y. acknowledges support from NSF grant PHY-1114374 and NASA grant NNX11AI49G, under sub-award 00001944, and also the NSF CAREER Award PHY-1250636.


\bibliography{master}
\end{document}